\documentclass[conference]{IEEEtran}
\IEEEoverridecommandlockouts
\usepackage{cite}
\usepackage{amsmath,amssymb,amsfonts}
\usepackage{algorithmic}
\usepackage{graphicx}
\usepackage{textcomp}
\usepackage{xcolor,colortbl}
\usepackage[caption=false,font=footnotesize]{subfig}
\usepackage{multirow}
\usepackage{hhline}
\usepackage{array}
\usepackage{transparent}
\usepackage{siunitx}
\usepackage{hyperref}

\definecolor{Gray}{gray}{0.8}

\graphicspath{{./figures/}}

\def\BibTeX{{\rm B\kern-.05em{\sc i\kern-.025em b}\kern-.08em
    T\kern-.1667em\lower.7ex\hbox{E}\kern-.125emX}}
    
\DeclareMathOperator*{\argmin}{arg\,min}

\DeclareMathOperator{\var}{Var}

\begin{document}

\title{Joint axis estimation for fast and slow movements using weighted gyroscope and acceleration constraints}

% Authors
\author{\IEEEauthorblockN{Fredrik~Olsson\IEEEauthorrefmark{1},~Thomas~Seel\IEEEauthorrefmark{2},~Dustin~Lehmann\IEEEauthorrefmark{3}~and~Kjartan~Halvorsen\IEEEauthorrefmark{4}} \IEEEauthorblockA{\IEEEauthorrefmark{1}\IEEEauthorrefmark{4}Department~of~Information~Technology, Uppsala University, Uppsala, Sweden} \IEEEauthorblockA{\IEEEauthorrefmark{2}\IEEEauthorrefmark{3}Department of Electrical Engineering and Computer Science, Technische Universit{\"a}t Berlin, Berlin, Germany} \IEEEauthorblockA{\IEEEauthorrefmark{4}Department of Mecatronics, Tecnol{\'o}gico de Monterrey, Mexico City, Mexico} \IEEEauthorblockA{Email: \IEEEauthorrefmark{1}fredrik.olsson@it.uu.se, \IEEEauthorrefmark{2}seel@control.tu-berlin.de,\\ \IEEEauthorrefmark{3}dustin.m.lehmann@campus.tu-berlin.de, \IEEEauthorrefmark{4}kjartan.halvorsen@it.uu.se} \thanks{This work was supported by the project \textit{Mobile assessment of human balance} (Contract number: 2015-05054), funded by the Swedish Research Council.}}

\maketitle

\begin{abstract}
% intro:
Sensor-to-segment calibration is a crucial step in inertial motion tracking. When two segments are connected by a hinge joint, for example in human knee and finger joints as well as in many robotic limbs, then the joint axis vector must be identified in the intrinsic sensor coordinate systems. There exist methods that identify these coordinates by solving an optimization problem that is based on kinematic joint constraints, which involve either the measured accelerations or the measured angular rates.
% methods:
In the current paper we demonstrate that using only one of these constraints leads to inaccurate estimates at either fast or slow motions. We propose a novel method based on a cost function that combines both constraints. The restrictive assumption of a homogeneous magnetic field is avoided by using only accelerometer and gyroscope readings. To combine the advantages of both sensor types, the residual weights are adjusted automatically based on the estimated signal variances and a nonlinear weighting of the acceleration norm difference. 
% results:
The method is evaluated using real data from nine different motions of an upper limb exoskeleton. Results show that, unlike previous approaches, the proposed method yields accurate joint axis estimation after only five seconds for all fast and slow motions.
\end{abstract}


\begin{figure}[t!]
\centering
\fontsize{17pt}{20pt}\selectfont%
\subfloat{\resizebox{80mm}{!}{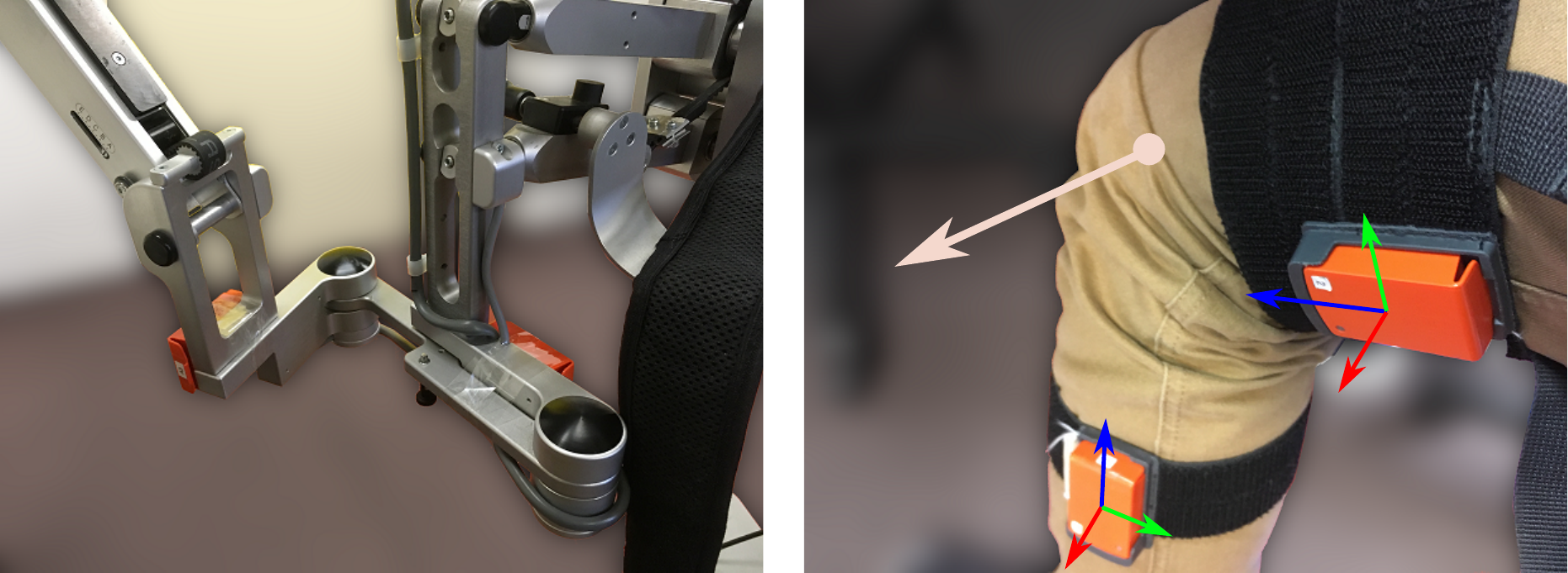}}%
\qquad
\vspace{.1cm}
\subfloat{\includegraphics[width=.8\linewidth]{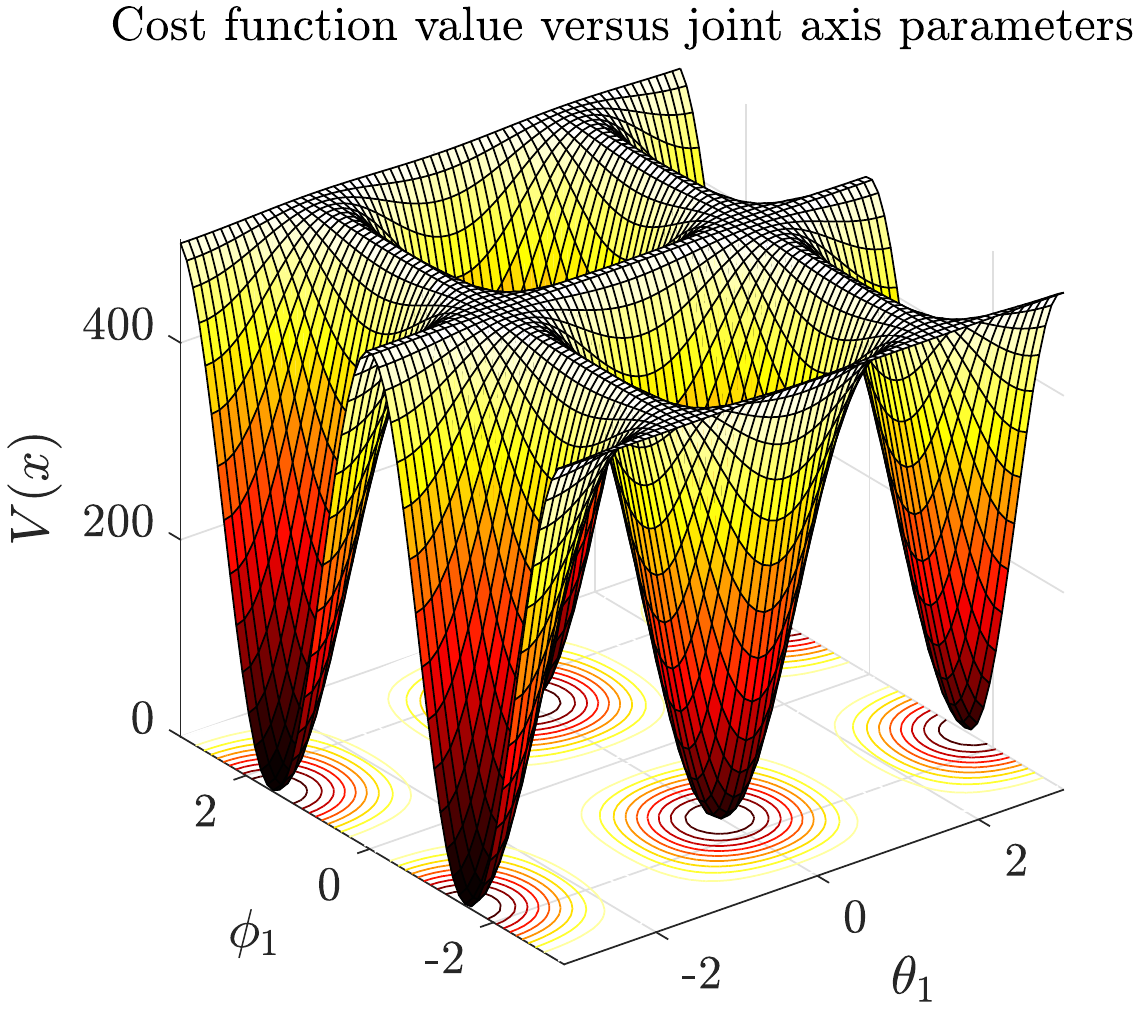}}
\caption{Top figures show two real-life examples of kinematic chains with a 1-DOF joint. Top left shows an exoskeleton frame and right shows a human knee joint. The joint axes are indicated by the arrows labeled $j$. Two 6-DOF IMUs are shown, with their three orthonormal sensor axes indicated by the arrows. The bottom figure illustrates the optimization-based approach to identify the joint axis. The joint axis is parametrized by the spherical coordinates $\theta_1$ and $\phi_1$, which are included in the unknown parameter vector $x$ to be estimated. We seek to minimize the cost function $V(x)$. The multiple local minima are all valid joint axes due to the parametrization and the fact that both $\pm j$ are valid joint axes.}
\label{fig:frontfig}
\end{figure}

\section{Introduction}
Microelectromechanical systems (MEMS) inertial measurement units (IMUs) are popular sensors, used to track movement in many different applications due to their mobility, low cost and small size. Areas where inertial sensors are used include, but are not limited to, medical rehabilitation, sports, robotics and navigation \cite{ahmad2013reviews}. A typical IMU measures 6 degrees of freedom (DOF), using a 3D accelerometer to measure linear acceleration (translational movement) and a 3D gyroscope to measure the rate of change in the sensor's orientation (rotational movement). In theory, a 6-DOF IMU measures all the quantities needed to fully track rigid body movement. However, since the IMU measurements must be integrated to obtain orientation and position estimates, any systematic error in the signal will accumulate. Orientation and position estimates are therefore typically only accurate for short periods of time, something which is often referred to as integration drift.

To alleviate the effects of integration drift, many different techniques have been proposed. Stationary accelerometer measurements can be used as absolute measurements of the inclination of the  sensor, i.e.~the angle of the sensor with respect to (w.r.t.) the global vertical axis. Using an additional 3D magnetometer sensor, which measures the local magnetic field, can give an absolute heading measurement, similar to a compass \cite{kok2017using}. A 6-DOF IMU combined with a 3D magnetometer is sometimes referred to as a 9-DOF IMU, as the magnetometer is typically soldered to the standard IMU chip. For position tracking, additional information can be derived from multiple measurements of ranges/distances to base stations at known locations. This includes Global positioning systems (GPS), and local ranging positioning systems, using for example ultrawideband ranging \cite{kok2015indoor}. In many applications however, adding additional hardware to compensate for the integration drift is not feasible, since the system then becomes too obtrusive or expensive. In such situations it becomes advantageous to make use of a model of the mechanical properties of the tracked object.

 Kinematic chains are often used to model human body segments or robotic manipulator systems. The kinematic chain consist of multiple rigid body segments connected by joints with 1 to 3 DOF. The DOF of a joint determines how many free variables (joint angles) that are required to fully describe the relative orientation of the adjacent segments. For kinematic chains, one can then exploit prior knowledge about the sensors' position w.r.t.~the joints and relative orientation w.r.t. the joint axes of 1-DOF and 2-DOF joints to simplify the tracking problem, or make it easier to combine the information from multiple IMUs attached to the various segments of the kinematic chain (see for example \cite{5504750} and \cite{Laidig2017_ICORR}).
 
 Such prior knowledge can of course be obtained by accurately measuring the placement of each sensor. However, a more user-friendly, less restrictive and less time-consuming solution is to allow the sensors to be placed arbitrarily and then use the available measurements to identify the sensor placement. Such methods have been proposed and used for the identification of the joint center of a 3-DOF joint \cite{SeelSchauerRaisch2012,CraboluPaniCereatti2016,olsson2017experimental} or the joint axis of a 1-DOF \cite{SeelSchauerRaisch2012,s18061882} and a 2-DOF joint \cite{laidig_2017_2d}. What these methods have in common is that they make use of the kinematic constraints of the underlying kinematic chain model to relate the measured accelerations and angular velocities to the unknown quantities to be identified, i.e.~they fit the kinematic model to the data. 

The current paper deals with the problem of estimating the direction of the axis of rotation in a joint connecting two segments. This joint axis estimation problem is further illustrated in Figure~\ref{fig:frontfig}. In addition to the usefulness of a kinematic model for tracking the motion of the mechanism, such estimation of the axis of rotation is important in positioning and adjusting exoskeletons \cite{jarrasse2012connecting} so that the axes of rotation of the exoskeleton and the human match.

We propose a novel method for joint axis estimation that overcomes limitations of previously proposed methods. The method in \cite{SeelSchauerRaisch2012} uses a gyroscope-based constraint, which requires that both segments rotate fast enough to let gyroscope bias become negligible. Moreover, even in the best case, the gyroscope-based constraint yields joint axis coordinates with ambiguous sign pairing \cite{Nowka2019_ECC}. Recently, a method that combines gyroscope and accelerometer readings has been proposed \cite{kuderle2018increasing}, however it requires stationary accelerometer measurements. We overcome this limitation by using a more general acceleration-based constraint that holds when the rotational and tangential acceleration components with respect to the joint axis are negligible. To balance between gyroscope and accelerometer information for movements of arbitrary speed, we use a weighted nonlinear optimization approach. More precisely, each data sample is weighted according to how much information about the joint axis it holds. This weighting approach allows accelerometer information to be included without requiring measurements under stationary conditions, and we will demonstrate that it provides accurate estimates with correct sign pairing for both fast and slow motions.

The remainder of the paper is structured as follows; The kinematic constraints of a hinge joint system are derived in Section~\ref{sec:kinematics}. In Section~\ref{sec:estimation} we present the nonlinear optimization method, which makes use of the weighted kinematic constraints to estimate the joint axes. The proposed method is evaluated using experimental data collected from an exoskeleton frame moved by a human. The experimental data, the evaluation method and metrics are explained in Section~\ref{sec:experiment}. Results are presented in Section~\ref{sec:results} and then discussed in Section~\ref{sec:discussions}. Finally, conclusions are drawn and potential future work outlined in Section~\ref{sec:conclusions}.

%% KINEMATIC MODEL %%
\section{Kinematic model} \label{sec:kinematics}
\subsection{Kinematic constraints of two segments in a a kinematic chain} \label{sec:kinematic_general}
Consider first the kinematic chain model where we have two rigid body segments connected by a joint. The joint can be either 1-, 2- or 3-DOF. Furthermore, we consider the case where each segment has one IMU rigidly attached to it in an arbitrary position and orientation. We define two Cartesian coordinate frames, referred to as the \textit{sensor frames}, $S_1$ and $S_2$, that are fixed in the center of the accelerometer triad of each IMU. We also define a global coordinate frame, $G$, which is fixed with respect to the local environment. We let subscripts $i \in \{1,2$\} denote quantities belonging to a specific sensor frame. We have that \begin{align}
a_i^{S_i} &= a_0^{S_i} + \omega_i^{S_i} \times (\omega_i^{S_i} \times r_i^{S_i}) + \dot{\omega}_i^{S_i} \times r_i^{S_1} \label{eq:a1}
\end{align} where $a_i$ are the accelerations of the sensor frames, $a_0$ is the acceleration of the joint center, $\omega_i$ and $\dot{\omega}_i$ are the angular velocities and angular accelerations of the sensor frames. The positions of the joint center with respect to each sensor frame are denoted by $r_i$, which we assume to be unknown. Each quantity in \eqref{eq:a1} are vectors in $\mathbb{R}^3$ since they explain arbitrary movements of a real-world object. The superscripts $S_i$ are used to indicate when a certain quantity is expressed in one of the two sensor frames. If we know the orientation of each sensor frame with respect to the global frame we can express the quantities in the global frame instead. For example \begin{align}
a_0^G &= R_1^{GS_1}a_0^{S_1} = R_2^{GS_2}a_0^{S_2},
\end{align} where $R_i^{GS_i}$ are the $3 \times 3$ rotation matrices that maps a vector expressed in $S_i$ into the global frame. The multiplication between a rotation matrix and a vector is equivalent to an orthonormal change of basis. 

For convenience we shall for the remainder of this document drop the use of the superscripts except for where it's needed. The relationship in \eqref{eq:a1} is linear in $a_0$ and $r_i$ and can equivalently be formulated as \begin{align}
a_i &= a_0^{S_i} + K(\omega_i,\dot{\omega}_i)r_i \label{eq:a1r}
\end{align} where \begin{align} \label{eq:kmat}
K(\omega,\dot{\omega}) = \begin{bmatrix}
-\omega_y^2 -\omega_z^2 & \omega_x \omega_y - \dot{\omega}_z & \omega_x \omega_z + \dot{\omega}_y \\
\omega_x \omega_y + \dot{\omega}_z  & -\omega_x^2 -\omega_z^2 & \omega_y \omega_z - \dot{\omega}_x \\
\omega_x \omega_z -\dot{\omega}_y &  \omega_y \omega_z + \dot{\omega}_x & -\omega_x^2 -\omega_y^2
\end{bmatrix},
\end{align} where subscripts $x,y,z$ denote the elements of the three-dimensional vectors. For convenience of notation we will write $K_i = K(\omega_i,\dot{\omega}_i)$.

\begin{figure}[!tb]
\centering{\footnotesize{%
\resizebox{70mm}{!}{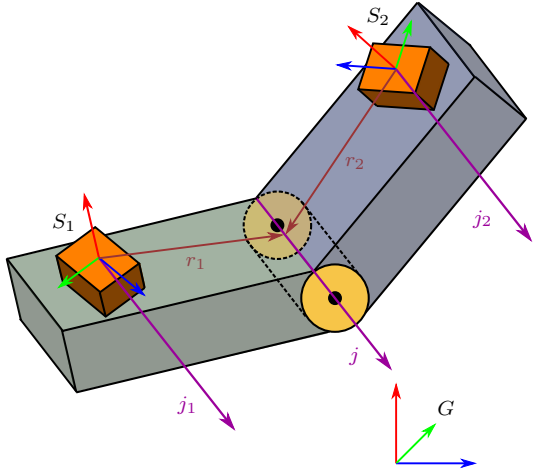}
}}%
\caption{The type of hinge joint system that we consider here. The two segments rotate independently with respect to each other only along the joint axis $j$. The sensor frames $S_i$ are rigidly fixed to their respective segments and their relative orientation can be described by one joint angle.}
\label{fig:hingeJoint}
\end{figure}

%\begin{figure}[!tb]
%    \centering
%    \includesvg[width=.8\linewidth]{hingeJoint}%
%    \caption{The type of hinge joint system that we consider here. The two segments rotate independently with respect to each other only along the joint axis $j$. The two sensor frames $S_1$ and $S_2$ are rigidly fixed to their respective segments and their relative orientation can be described by one joint angle.}
%    \label{fig:hingeJoint}
%\end{figure}

%% HINGE JOINT KINEMATICS %%
\subsection{Kinematic constraints of a hinge joint system}  \label{sec:kinematics_hinge}
For a 1-DOF joint, the two segments can only rotate independently with respect to each other along the \textit{joint axis}. We let $\|\cdot \|$ denote the Euclidean vector norm, then the joint axis is defined by the unit vector $j \in \mathbb{R}^3, \| j \| = 1$.  We refer to such a joint as a \textit{hinge joint}.
From its definition, the joint axis vector, $j$, must satisfy \begin{align} \label{eq:joint_axis_property}
R_1 j_1 &= R_2 j_2,
\end{align} meaning that the joint axes $j_i$ expressed in the two sensor frames coincide in the global frame, see Figure~\ref{fig:hingeJoint}. We can decompose the angular velocities into one component that is parallel to the joint axis and one that is perpendicular to the joint axis \begin{align}
\omega_i &= \omega_{j_i} + \omega_{j_i^\perp} \\
\omega_{j_i} &= j_i^\top \omega_i j_i \\
\omega_{j_i^\perp} &= \omega_i - \omega_{j_i} = \omega_i - j_i^\top \omega_i j_i.
\end{align} Since the two segments can only rotate independently along the joint axis it follows that the perpendicular components must have the same magnitude regardless of reference frame \begin{align} \label{eq:angvel_perp}
\| \omega_{j_1^\perp} \| &= \| \omega_{j_2^\perp} \|.
\end{align} The magnitude of the perpendicular component can also be computed from the cross product between the angular velocity and the joint axis \begin{align} \label{eq:angvel_perp_mag}
\| \omega_i - j_i^\top \omega_i j_i \| &= \| \omega_i \times j_i \|.
\end{align} Combining \eqref{eq:angvel_perp} and \eqref{eq:angvel_perp_mag} we formulate the \textit{angular velocity constraint} \begin{align} \label{eq:angvel_constraint}
\|\omega_1 \times j_1 \| - \| \omega_2 \times j_2 \| = 0,
\end{align} which must be satisfied by hinge joint systems.

If we now look at the projection of the accelerations onto the joint axis, from \eqref{eq:a1r} we have that \begin{align}
j_i^\top a_i j_i &= j_i^\top a_0^{S_i} j_i + j_i^\top K_i r_i j_i.    
\end{align} It then follows from \eqref{eq:joint_axis_property} that \begin{align} \begin{split}
   R_1 j_1 j_1^\top a_0^{S_1} &= R_2 j_2 j_2^\top a_0^{S_2} \Longrightarrow \\
j_1^\top a_0^{S_1} &= j_2^\top a_0^{S_2}. 
\end{split}
\end{align} By projecting the accelerations onto the joint axis and subtracting one from the other we get \begin{align} \label{eq:acc_constrant_nonzero} \begin{split}
  j_1^\top a_1 - j_2^\top a_2 &= j_1^\top a_0^{S_1} - j_2^\top a_0^{S_2}+ j_1^\top K_1 r_1 - j_2^\top K_2 r_2 \\
&=  j_1^\top K_1 r_1 - j_2^\top K_2 r_2,  
\end{split}
\end{align} where we see that only the rotational components of the accelerations remain on the right hand side. It's clear that if the radial and tangential acceleration components along the direction of the joint axis are small ($j_i^\top K_i r_i \approx 0, \, \forall{i}$) the right hand side will vanish \begin{align} \label{eq:acc_constraint}
j_1^\top a_1 - j_2^\top a_2 &\approx 0,
\end{align} which forms the \textit{acceleration constraint} for the hinge joint system.

%% ESTIMATION %%
\section{Estimation method} \label{sec:estimation}
\subsection{Nonlinear optimization}
Since the joint axes $j_i$ are unit vectors, we parametrize them using spherical coordinates to implicitly enforce the unit vector constraint \begin{align}
x &= \begin{bmatrix}
\theta_1 & \phi_1 & \theta_2 & \phi_2
\end{bmatrix}^\top \\
j_{i}(x) &= \begin{bmatrix}
\cos \theta_i \cos \phi_i \\
\cos \theta_i \sin \phi_i \\
\sin \theta_i
\end{bmatrix},
\end{align} which then become the unknown parameters to estimate. The estimation problem can then be formulated as \begin{align} \label{eq:optim_joint_axis}
\hat{x} &= \argmin_x V(x) \\
V(x) &= \sum_{k=1}^N (e_\omega(k,x))^2 + (e_a(k,x))^2,
\end{align} where $e_\omega(k,x)$ and $e_a(k,x)$ are residual terms, based on the angular velocity- and acceleration constraints given by \eqref{eq:angvel_constraint} and \eqref{eq:acc_constraint} \begin{align}
e_\omega(k,x) &= w_\omega(k)(\|\omega_1(t_k) \times j_1(x) \| - \|  \omega_2(t_k) \times j_2(x) \|) \label{eq:gyro_residual} \\
e_a(k,x) &= w_a(k)(j_1^\top a_1(t_k) - j_2^\top a_2(t_k)) \label{eq:acc_residual}.
\end{align} Important to note here is that we now use the IMU measurements of the kinematic variables $\omega_i$ and $a_i$, where $t_k, \, k=1,\ldots,N$ denotes the sampling time of the $k$:th data sample used for estimation. The two scalars $w_\omega(k)$ and $w_a(k)$ are weights that are used to give more or less importance to specific samples. A strategy for selecting these weights is proposed in Section~\ref{sec:estimation_weights}.

The optimization problem \eqref{eq:optim_joint_axis} is a minimization problem for a sum of squared residual terms. An efficient method for solving such problems is the Gauss-Newton method \cite{nocedalopt}, which uses the gradients of the residuals to estimate the Hessian of the cost function, which is then used to select the search direction of the optimization method.

The gradients of the residuals are computed in the following way \begin{align}
\frac{\partial e_\omega(k,x)}{\partial x} &= \frac{\partial j}{\partial x} \frac{\partial e_\omega(k,x)}{\partial j}w_\omega(k) \end{align} \begin{align}
\frac{\partial e_\omega(k,x)}{\partial j} &= \begin{bmatrix}
\frac{\partial (\|\omega_1(t_k) \times j_1(x) \| ) }{\partial j_1} \\
- \frac{\partial (\|\omega_2(t_k) \times j_2(x) \| ) }{\partial j_2}
\end{bmatrix} \end{align} \begin{align}
\frac{\partial (\|\omega_i(t_k) \times j_i(x) \| ) }{\partial j_i} &= \frac{(\omega_i(t_k) \times j_i) \times \omega_i(t_k)}{\|\omega_i(t_k) \times j_i(x) \|} \\
\frac{\partial e_a(k,x)}{\partial x} &= \frac{\partial j}{\partial x} \frac{\partial e_a(k,x)}{\partial j}  \end{align} \begin{align}
\frac{\partial e_a(k,x)}{\partial j} &= \begin{bmatrix}
a_1(t_k) \\ -a_2(t_k)
\end{bmatrix}w_a(k) \end{align} \begin{align}
\frac{\partial j}{\partial x} &= \begin{bmatrix}
\frac{\partial j_1}{\partial x_1} & 0 \\
0 & \frac{\partial j_2}{\partial x_2}
\end{bmatrix} \end{align} \begin{align}
\frac{\partial j_i}{\partial x_i} &= \begin{bmatrix}
-\sin \theta_i \cos \phi_i & -\cos \theta_i \sin \phi_i \\
-\sin \theta_i \sin \phi_i & \cos \theta_i \cos \phi_i \\
 \cos \theta_i &  0
\end{bmatrix}^\top.
\end{align} 

\subsection{Weighting the residuals} \label{sec:estimation_weights}
The weights $w_\omega(k)$ and $w_a(k)$ that appear in the residuals \eqref{eq:gyro_residual}--\eqref{eq:acc_residual} should in some sense reflect the amount of information the $k$:th sample contain about the joint axis. In this section we propose a strategy for selecting these weights.

The first thing to consider is that the two residuals \eqref{eq:angvel_constraint} and \eqref{eq:acc_constraint} contain measurements of two different physical quantities, angular velocity [\SI{}{\radian\per\second}] and acceleration [\SI{}{\metre\per\second\squared}]. Because of this, we expect the variance of the residuals to have different magnitudes. To estimate the variance of the residuals we assume that the gyroscope and accelerometer measurements are random variables that belong to multivariate Gaussian distributions \begin{align}
\omega(t_k) &= \begin{bmatrix} \omega_1^\top(t_k) & \omega_2^\top(t_k)
\end{bmatrix}^\top \\
\omega(t_k) &\sim \mathcal{N}(\mu_\omega(t_k),\Sigma_\omega) \\
a(t_k) &= \begin{bmatrix} a_1^\top(t_k) & a_2^\top(t_k)
\end{bmatrix}^\top \\
a(t_k) &\sim \mathcal{N}(\mu_a(t_k),\Sigma_a),
\end{align} where $\mu_{\omega,a}$ are the means, which depend on the movement that each sensor is undergoing at time $t_k$ and the $\Sigma_{\omega,a}$ are covariance matrices, assumed to be diagonal and time-independent. The means are considered unknown since they depend on the instantaneous angular velocities and accelerations of the sensors. The covariance matrices, however, can be estimated as the sample covariance of a stationary sensor. Which is typically done as part of the IMU calibration.

Because the gyroscope residual is nonlinear, we approximate its variance using the delta-method, which is based on a first-order Taylor approximation of a nonlinear function of a random variable around its mean value \begin{align}
\var \{ e_\omega(k,x) \} &\approx \left( \frac{\partial e_\omega(k,x)}{\partial \omega(t_k)} \right)^\top \Sigma_\omega \frac{\partial e_\omega(k,x)}{\partial \omega(t_k)}  \\
\frac{\partial e_\omega(k,x)}{\partial \omega(t_k)} &= \begin{bmatrix}
\frac{(j_1 \times \omega_1(t_k)) \times j_1}{\|\omega_1(t_k) \times j_1(x) \|} \\
\frac{(j_2 \times \omega_2(t_k)) \times j_2}{\|\omega_2(t_k) \times j_2(x) \|}
\end{bmatrix}. \label{eq:gyr_resid_diff}
\end{align} Important to note here is that the partial derivatives should be evaluated at the mean $\mu_\omega(t_k)$, but this is not known since it depends on the true angular velocity of the system. However, the partial derivatives in \eqref{eq:gyr_resid_diff} are unit vectors since \begin{align}
    \| (j_i \times \omega_i(t_k)) \times j_i \| &= \| \omega_i(t_k) \times j_i(x) \|.
\end{align} If we then choose to approximate the covariance matrix of the measurements as \begin{align}
    \Sigma_{\omega,a} &\approx \sigma_{\omega,a}^2 I \\
    \sigma_{\omega,a}^2 &= \max \{ \var \{\omega(t_k),a(t_k) \} \},
\end{align} where $I$ is the unit matrix and $\sigma_{\omega,a}^2$ is the maximum sample variance among all sensor axes. We then get a worst-case estimate of the residual variance, which is independent of the unknown sensor axes \begin{align} \begin{split} \var \{ e_\omega(k,x) \} &\approx \sigma_\omega^2 \left( \frac{\partial e_\omega(k,x)}{\partial \omega(t_k)} \right)^\top I \frac{\partial e_\omega(k,x)}{\partial \omega(t_k)}  \\
&= 2 \sigma_\omega^2. \end{split}
\end{align} The variance of the accelerometer residual is easier to compute as it is linear. If we use the same worst-case variance approximation as for the gyroscope residual, we get \begin{align} \begin{split}
\var \{ e_a(k,x) \} &= \begin{bmatrix} j_1^\top & -j_2^\top \end{bmatrix} \Sigma_a \begin{bmatrix} j_1 \\ -j_2 \end{bmatrix} \\
&\approx \sigma_a^2 \begin{bmatrix} j_1^\top & -j_2^\top \end{bmatrix} I \begin{bmatrix} j_1 \\ -j_2 \end{bmatrix} \\
&= 2\sigma_a^2. \end{split}
\end{align}

We use the ratio of the estimated variances to select a base weight, $w_0$ as \begin{align}
    w_0 &= \sqrt{\frac{\var  \{ e_a \}}{\var  \{ e_\omega \}}} = \sqrt{\frac{2\sigma_a^2}{2\sigma_\omega^2}} = \frac{\sigma_a}{\sigma_\omega},
\end{align} which is equal to the ratio of the worst-case sample standard deviations of the accelerometers and gyroscopes. Now if we select $w_\omega(k) = w_0$, the variances of the residuals will have similar magnitudes when both kinematic constraints are valid. 

However, as previously shown, the acceleration constraint implies  $j_1^\top K_1 r_1 \approx 0$ and $j_2^\top K_2 r_2 \approx 0$. Since we do not assume any prior knowledge of $j_i$ nor $r_i$ we will use the difference in magnitude of the two accelerometer measurements as a penalty term to lower the accelerometer residual weights. We use the fact that \begin{align}
    K_1 r_1 - K_2 r_2 = 0 \Rightarrow \| a_1 \| - \| a_2 \| = 0, \label{eq:acc_satisfied}
\end{align} to select the accelerometer residual weight as \begin{align} \label{eq:acc_weight}
    w_a(k) &= \sqrt{\frac{1}{1+(\| a_1(t_k) \| - \| a_2(t_k) \|)^2}}.
\end{align} An interpretation of this weighting strategy is that we choose $(\| a_1(t_k) \| - \| a_2(t_k) \|)^2$ as the pseudo-variance of an error term that invalidates the acceleration constraint. Note that accelerations that have equal linear and rotational components will also yield $w_a(k) = 1$ even though they violate the constraint. However, when \eqref{eq:acc_satisfied} is not satisfied, we know that the acceleration constraint is also not satisfied. An alternative strategy could be to select the error term based on $K_1 r_1 - K_2 r_2$, but since the rotational acceleration components depend on the unknown joint center positions $r_i$, we deemed the chosen $w_a(k)$ to be less restrictive.

\begin{figure}[tb!]
\centering
\subfloat{\includegraphics[width=0.43\linewidth]{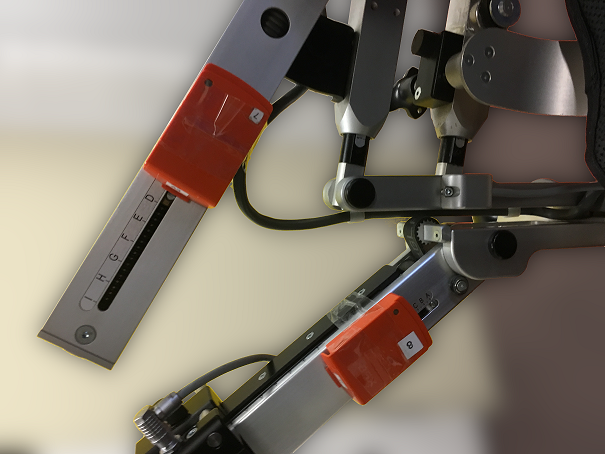}}
\quad
\subfloat{\includegraphics[width=0.43\linewidth]{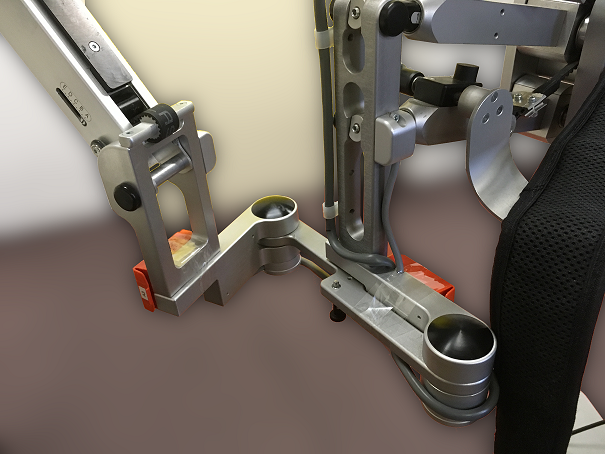}}
\caption{The experimental setup with the two IMUs attached to the exoskeleton frame, where the joint axis is horizontal (left) and vertical (right).}
\label{fig:experiment}
\end{figure}

\section{Experimental evaluation} \label{sec:experiment}
\subsection{Data acquisition}
Data were collected from two Xsens wireless IMUs \cite{Xsens2017} attached to different segments of an upper limb exoskeleton frame that were connected by a hinge joint, see Figure~\ref{fig:experiment}. The exoskeleton was controlled by a human, performing motions where both segments move and the joint undergoes repeated flexion and extension at varying speed. First, a completely arbitrary motion is performed, where the entire joint rotated freely in 3D space. Then the joint moves such that the joint axis remains vertical while both segments move in the same horizontal plane. Finally, that planar motion is performed again but with a horizontal joint axis. The first motion is chosen to show that the methods work for arbitrary motions, while the second is chosen to show that it is not required to move the joint around all degrees of freedom. The third motion is included because recent theoretical analysis shows that for such motions the joint axis coordinates are only identifiable up to sign pairing \cite{Nowka2019_ECC}. A total of nine data sets were collected: \begin{itemize}
    \item Freely moving joint axis: Fast, slow and mixed movement speed
    \item Vertical joint axis: Fast, slow and mixed movement speed
    \item Horizontal joint axis: Fast, slow and mixed movement speed
\end{itemize} %We characterize the fast, slow and mixed movement speed by the empirical cumulative distribution function (CDF) of the measured angular velocity magnitudes, see Figure~\ref{fig:speed}.
The fast data has angular velocity magnitudes up to \SI{4.1}{\radian\per\second} (Mean $=\SI{1.3}{\radian\per\second}$, SD $=\SI{0.9}{\radian\per\second}$) and the slow data up to \SI{0.6}{\radian\per\second} (Mean $=\SI{0.09}{\radian\per\second}$, SD $=\SI{0.08}{\radian\per\second}$). The mixed data alternates between fast and slow movement in intervals of \SIrange{5}{10}{\second}.

Before each data set we collected \SI{10}{\second} of measurements where the sensors were stationary. This stationary data was used to calibrate the IMUs. A bias parameter was estimated as the sample mean of the stationary gyroscope measurements. This constant bias was then subtracted from each measurement of the respective following data sets. For the accelerometer a scalar gain parameter was estimated using the least squares method in order to make the average magnitude of the stationary accelerometer measurements match that of the gravitational acceleration, $g = \SI{9.81}{\metre\per\second\squared}$. Each accelerometer measurement was then multiplied by this gain parameter.

\begin{table*}[tb!]
\renewcommand{\arraystretch}{1.3}
\caption{MAD and SAD metrics from the methods using the non-weighted residuals, $e_\omega$ and $e_a$ by themselves and combined without and with the proposed weighting variables. The lowest MAD values and the MAD values within \ang{0.5} are highlighted in bold for each axis--movement speed combination.}
\label{table:results}
\centering
\begin{tabular}{cc|c|c| |c|c| |c|c|}
\cline{3-8}
& & \multicolumn{2}{ c|| }{Free axis} & \multicolumn{2}{ c|| }{Vertical axis} & \multicolumn{2}{ c| }{Horizontal axis} \\ \cline{3-8}
& & \multicolumn{1}{ c| }{$\hat{j}_1$} & \multicolumn{1}{ c|| }{$\hat{j}_2$} & \multicolumn{1}{ c| }{$\hat{j}_1$} & \multicolumn{1}{ c|| }{$\hat{j}_2$} & \multicolumn{1}{ c| }{$\hat{j}_1$} & \multicolumn{1}{ c| }{$\hat{j}_2$} \\ \cline{1-8}
\multicolumn{1}{|c}{Speed} & \multicolumn{1}{|c||}{Residuals} & \multicolumn{1}{ c| }{MAD(SAD) [$^\circ$]} & MAD(SAD) [$^\circ$] & \multicolumn{1}{ c| }{ MAD(SAD) [$^\circ$]} & MAD(SAD) [$^\circ$] &  \multicolumn{1}{ c| }{MAD(SAD) [$^\circ$]} & MAD(SAD) [$^\circ$] \\ \hhline{|=|=||=|=||=|=||=|=|}
%Fast movement block
\multicolumn{1}{ |c  }{\multirow{4}{*}{Fast} } &
\multicolumn{1}{ |c|| }{$e_\omega$} & {\cellcolor{Gray} 2.2(1.6)} & {\cellcolor{Gray} 81.9(87.4)} & 1.4(1.7) & 90.6(87.2) & {\cellcolor{Gray} \textbf{8.2}(5.4)} & {\cellcolor{Gray} 90.3(85.4)} \\ \hhline{|~|-||-|-||-|-||-|-|}
\multicolumn{1}{ |c  }{}                        &
\multicolumn{1}{ |c|| }{$e_a$} & {\cellcolor{Gray} 5.9(3.7)} & {\cellcolor{Gray} 5.9(3.3)} & 9.7(6.7) & 10.3(6.5) & {\cellcolor{Gray} 50.6(47.3)} & {\cellcolor{Gray} 63.4(52.0)} \\ \hhline{|~|-||-|-||-|-||-|-|}
\multicolumn{1}{ |c  }{}                        &
\multicolumn{1}{ |c|| }{$e_\omega + e_a$} & {\cellcolor{Gray} 7.6(6.7)} & {\cellcolor{Gray} 7.6(6.8)} & 4.9(3.7) & 5.2(3.7) & {\cellcolor{Gray} 31.6(31.2)} & {\cellcolor{Gray} 70.9(74.0)} \\ \hhline{|~|-||-|-||-|-||-|-|}
\multicolumn{1}{ |c  }{}                        &
\multicolumn{1}{ |c|| }{Weighted} & {\cellcolor{Gray} \textbf{0.7}(0.4)} & {\cellcolor{Gray} \textbf{0.5}(0.3)} & \textbf{0.7}(0.8) & \textbf{1.5}(1.3) & {\cellcolor{Gray} \textbf{8.3}(7.4)} & {\cellcolor{Gray} 90.8(85.7)} \\ \hhline{|=|=||=|=||=|=||=|=|}
%Slow movement block
\multicolumn{1}{ |c  }{\multirow{4}{*}{Slow} } &
\multicolumn{1}{ |c|| }{$e_\omega$} & \textbf{0.8}(0.4) & 75.2(87.1) & {\cellcolor{Gray} 2.0(1.2)} & {\cellcolor{Gray} 68.6(85.8)} & 3.4(3.0) & 90.9(88.5) \\ \hhline{|~|-||-|-||-|-||-|-|}
\multicolumn{1}{ |c  }{}                        &
\multicolumn{1}{ |c|| }{$e_a$} & 6.7(13.4) & 6.5(14.2) & {\cellcolor{Gray} 6.3(7.6)} & {\cellcolor{Gray} 6.9(7.4)} & 51.5(46.6) & 58.3(47.5) \\ \hhline{|~|-||-|-||-|-||-|-|}
\multicolumn{1}{ |c  }{}                        &
\multicolumn{1}{ |c|| }{$e_\omega + e_a$} & 3.3(11.5) & 4.1(12.2) & {\cellcolor{Gray} 4.4(2.4)} & {\cellcolor{Gray} 3.1(1.9)} & 48.2(35.0) & 77.9(52.1) \\ \hhline{|~|-||-|-||-|-||-|-|}
\multicolumn{1}{ |c  }{}                        &
\multicolumn{1}{ |c|| }{Weighted} & \textbf{0.7}(0.4) & \textbf{1.4}(1.0) & {\cellcolor{Gray} \textbf{1.0}(0.5)} & {\cellcolor{Gray} \textbf{1.6}(1.0)} & \textbf{2.4}(1.9) & 90.9(88.5) \\ \hhline{|=|=||=|=||=|=||=|=|}
%Mixed movement block
\multicolumn{1}{ |c  }{\multirow{4}{*}{Mixed} } &
\multicolumn{1}{ |c|| }{$e_\omega$} & {\cellcolor{Gray} \textbf{1.4}(1.0)} & {\cellcolor{Gray} 76.7(87.0)} & 2.5(1.8) & 79.8(84.6) & {\cellcolor{Gray} \textbf{5.2}(3.1)} & {\cellcolor{Gray} 90.6(86.9)} \\ \hhline{|~|-||-|-||-|-||-|-|}
\multicolumn{1}{ |c  }{}                        &
\multicolumn{1}{ |c|| }{$e_a$} & {\cellcolor{Gray} 26.2(17.4)} & {\cellcolor{Gray} 24.8(16.1)} & 17.7(11.5) & 18.0(12.1) & {\cellcolor{Gray} 39.4(24.9)} & {\cellcolor{Gray} 73.3(53.0)} \\ \hhline{|~|-||-|-||-|-||-|-|}
\multicolumn{1}{ |c  }{}                        &
\multicolumn{1}{ |c|| }{$e_\omega + e_a$} & {\cellcolor{Gray} 10.0(6.2)} & {\cellcolor{Gray} 7.8(5.4)} & 7.6(5.3) & 9.5(6.5) & {\cellcolor{Gray} 27.3(22.9)} & {\cellcolor{Gray} 82.1(70.0)} \\ \hhline{|~|-||-|-||-|-||-|-|}
\multicolumn{1}{ |c  }{}                        &
\multicolumn{1}{ |c|| }{Weighted} & {\cellcolor{Gray} \textbf{0.9}(0.6)} & {\cellcolor{Gray} \textbf{1.6}(1.2)} & \textbf{0.9}(0.8) & \textbf{2.1}(1.5) & {\cellcolor{Gray} \textbf{4.8}(3.0)} & {\cellcolor{Gray} 90.7(87.3)} \\ \hhline{|-|-||-|-||-|-||-|-|}
\end{tabular}
\end{table*}

\subsection{Evaluation method and metrics}
To evaluate the use of the proposed residual weighting method with the non-weighted alternatives we perform the following steps for each specific data set: \begin{enumerate}
    \item Divide the data set into $M$ segments spread out evenly over the length of the entire data set, where each segment contains $N$ time-consecutive measurements, i.e. $t_1 < t_2 < \ldots < t_N$.
    \item For each data segment, one random initial estimate $\hat{x}_0$ is sampled such that the initial estimates of $j_i$ are distributed uniformly on the unit sphere in $\mathbb{R}^3$.
    \item Compute the joint axis estimates $\hat{j}_i$ using all of the different methods to be evaluated.
    \item After all $M$ estimation runs have finished, compute evaluation metrics.
\end{enumerate} The metrics used for evaluation are the mean angular deviation (MAD) and standard angular deviations (SAD) \cite{kuderle2018increasing}. The angular deviation (AD) is defined as the angle between two estimated joint axis vectors \begin{align}
\text{AD}(\hat{j}_{i,m},\hat{j}_{i,n}) &= \cos^{-1}(\hat{j}_{i,m}^\top \hat{j}_{i,n}),
\end{align} where subscripts $m,n \in \{1,\ldots,M \}$ are used to denote specific estimates in the set of all estimated joint axes. The AD is then used to compute the MAD and SAD for $\hat{j}_i$ for each method to be evaluated. The MAD is computed as the sample mean of all different ADs in the set of estimated joint axes \begin{align}
    \text{MAD}(\hat{j}_i) &= \frac{\sum_{m=1}^M \sum_{n=1}^{m-1} \text{AD}(\hat{j}_{i,m},\hat{j}_{i,n})}{{M \choose 2}},
\end{align} and the SAD is computed as the sample standard deviation of the same set of ADs.
%\begin{align}
%    \text{SAD}(\hat{j}_i) &= \sqrt{ \frac{ \sum_{m=1}^M \sum_{n=1}^{m-1} (\text{AD}(\hat{j}_{i,m},\hat{j}_{i,n})-\text{MAD}(\hat{j}_i))^2} {{N \choose 2}-1}}.
%\end{align} 
Since we use a rigid mechanical exoskeleton frame with orthogonal edges for evaluation it is possible to place the sensors such that we have fairly accurate knowledge about the true joint axes $j_i$. The sensors were placed such that one of the edges of the sensor casing was as closely aligned with the joint axis as possible. Since the edges of the sensor casing are approximately aligned with the true sensor axes, the true joint axis should have an element close to $\pm 1$ at the element corresponding to the aligned sensor axis and the other two elements should be close to zero. Using this we compute the \textit{angular error} as $\text{AD}(\hat{j}_{i,m},j_i)$ and compute the means and standard deviations as a metric for how consistent the estimates are with respect to the true joint axis. 

Note that any joint axis can be described equally by a vector $j$ and its negative counterpart $-j$. Hence we consider both $(j_1, j_2)$ and $(-j_1,-j_2)$ as valid joint axis coordinates (correct sign pairing). However, it is important to note that $(-j_1,j_2)$, $(j_1,-j_2)$ are non-valid estimates (wrong sign pairing), as also discussed in \cite{Nowka2019_ECC}. For example, joint axis coordinates with wrong sign pairing lead to completely wrong results when used for calculating joint angles \cite{Laidig2017_ICORR}.

The AD metric should reflect the correct identification of the axis direction as well as the correct sign pairing. To assure this, we first compute $\text{AD}(\pm \hat{j}_{1,m},j_1)$ (with respect to the true joint axis) and select the sign which yields the lowest AD. If we then select a negative sign for $\hat{j}_{1,m}$ we do the same for $\hat{j}_{2,m}$ to maintain the original sign pairing. Doing this will cause the estimates of $j_1$ to be as consistent as possible, whereas the estimates for $j_2$ will only be consistent if the sign pairing is correct.

\section{Results} \label{sec:results}
The evaluations were performed for all $9$ different data sets, where each data set was divided into $M = 100$ data segments with $N = 500$ consecutive measurements in each segment. For the IMUs we used, this corresponded to \SI{5}{\second} of consecutive measured motion. The methods being compared use the gyroscope and accelerometer residuals alone, combined without weighting and combined with the weighting proposed in Section~\ref{sec:estimation_weights}. The resulting MAD and SAD values are summarized for all methods and data sets in Table~\ref{table:results}. The angular error metrics, which compare the joint axis estimates to the joint axis to which the sensors were aligned, are summarized in Figure~\ref{fig:ground_truth}.

\section{Discussion} \label{sec:discussions}
The compared methods were evaluated by their achieved MAD/SAD and angular errors. Lower MAD/SAD values indicate that the estimates are more consistent across the entire data set whereas a lower angular error indicates that the estimates are consistent with the joint axis to which the sensors were aligned. Even though the edges of the sensor casing was carefully aligned with the edges of the exoskeleton frame and the joint axis, we still expect some uncertainty in the alignment. Partially due to the placement itself and partially due to the fact that the true sensor axes may not be perfectly aligned with the sensor casing. A conservative estimate of this uncertainty is expected to be around \ang{5}. Therefore, we cannot draw conclusions based solely on angular errors if their magnitudes are close to that of the uncertainty. Angular errors larger than \ang{20} were deemed too inconsistent with the sensor placement, which is why Figure~\ref{fig:ground_truth} only shows errors less than that.

\subsection{Performance evaluation of the compared methods}
\begin{figure}[tb!]
\centering
\subfloat{\includegraphics[width=0.9\linewidth]{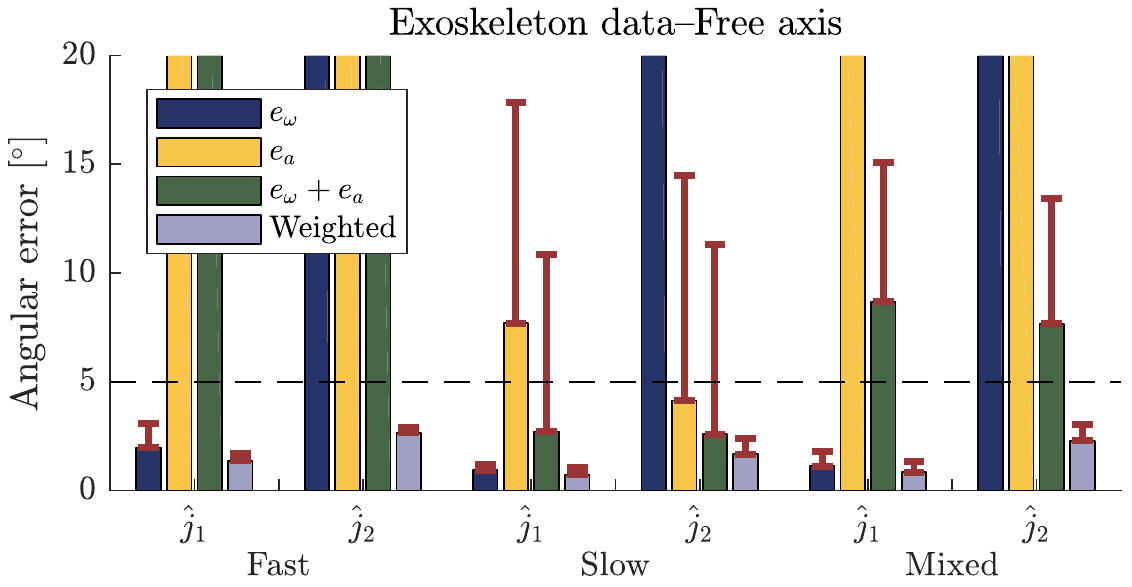}}
\qquad
\subfloat{\includegraphics[width=0.9\linewidth]{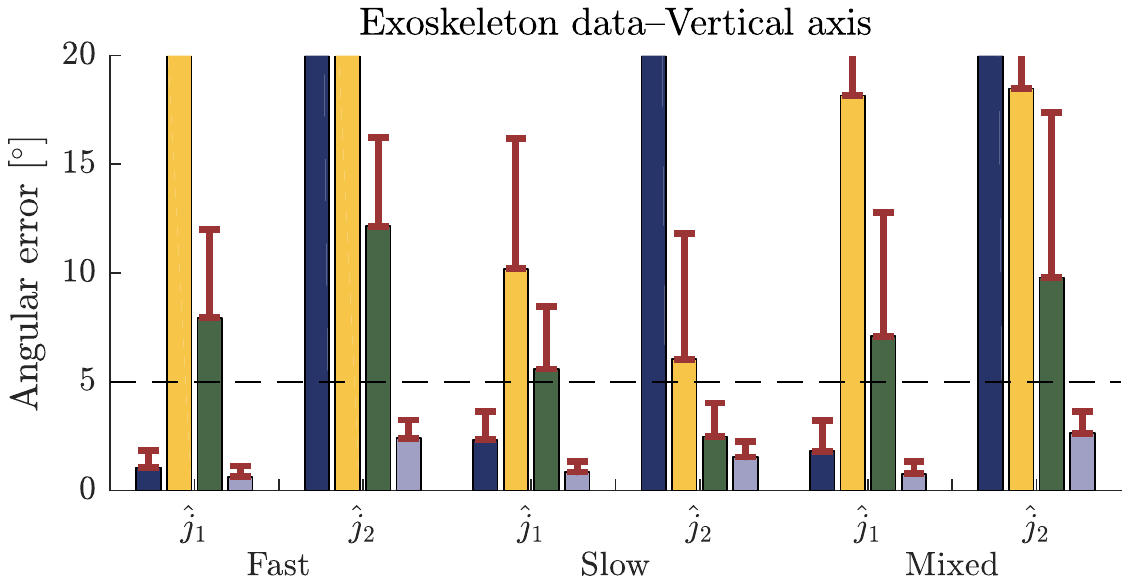}}
\qquad
\subfloat{\includegraphics[width=0.9\linewidth]{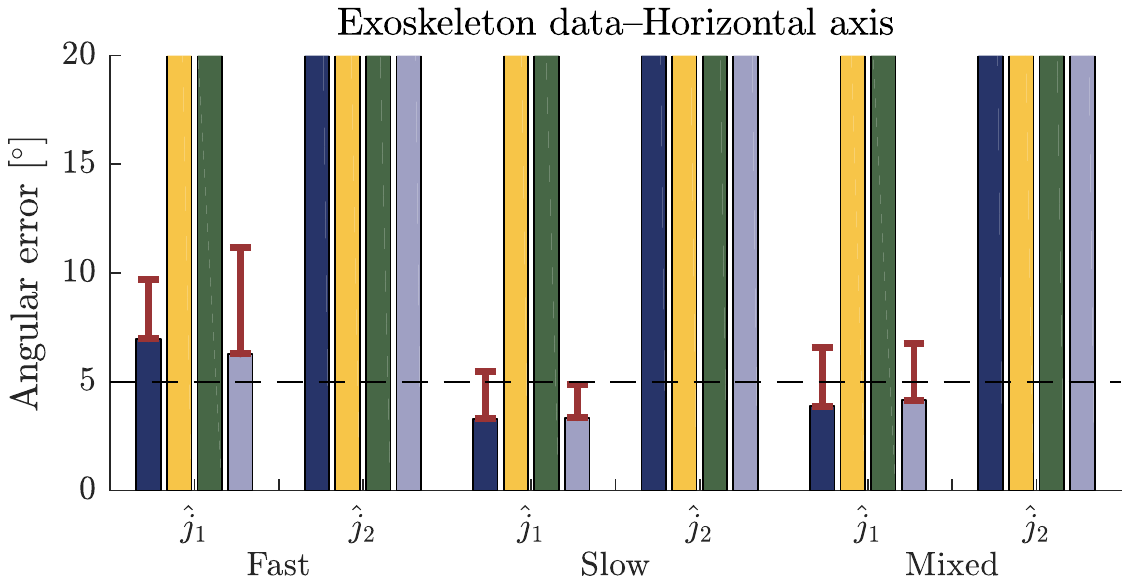}}
\caption{Angular errors computed as the angular deviation of the estimated joint axis w.r.t. the axis to which the sensors were aligned. From top to bottom: Free, vertical and horizontal axis. From left to right: Fast, slow and mixed movement speed. The mean angular errors for the compared methods are shown by the colored bars. Standard deviations are shown by the red error bars. The horizontal dashed line indicates a \ang{5} estimated uncertainty in the sensor alignment. The vertical axis only shows errors below \ang{20} as larger errors were deemed too inconsistent with the sensor placement.}
\label{fig:ground_truth}
\end{figure}

We first compare the results for the free and vertical axis data. The proposed method with the weighted gyroscope and accelerometer residuals had the best performance out of all compared methods. This is indicated by the bold faced MAD values in Table~\ref{table:results} and also by the lowest angular errors in Figure~\ref{fig:ground_truth}. This shows that the proposed method is both the most consistent method which also achieves correct sign pairing for the free and vertical axis data.

The gyroscope-only method performed similarly to the proposed method for $\hat{j}_1$, which were chosen as the minimizer of $AD(\pm \hat{j}_1,j_1)$. However, because the sign pairing is ambiguous for the gyroscope-only method \cite{Nowka2019_ECC} we see very large MAD/SAD and angular errors for $\hat{j}_2$. The metrics for $\hat{j}_1$ demonstrate the ideal performance of the method when we have prior knowledge about the sign pairing of the joint axes, for example by restricting the sensor attachment to a known surface, and the sensor facing a predetermined side of the segment.

The accelerometer-only and the combined non-weighted methods performed significantly worse in terms of MAD/SAD and angular errors, but achieves the correct sign pairing as indicated by the lower angular errors for $\hat{j}_2$ for the slow movement. Interesting to note that in some cases, these methods yielded relatively low MAD/SAD values, for examples the Free axis--Fast data in Table~\ref{table:results}. However, the angular errors for the same data were significantly larger as shown in Figure~\ref{fig:ground_truth}. This indicates that these methods have larger bias for fast and mixed movement speeds, which can be explained by the fact that the approximation used to define the acceleration constraint in \eqref{eq:acc_constrant_nonzero}--\eqref{eq:acc_constraint} is not valid for fast rotations along the joint axis. Combining the accelerometer and gyroscope yielded better results than accelerometer-only method, but the non-weighted combination was significantly worse than the weighted combination.

For the horizontal axis data we see that the proposed method and the gyroscope-only method performed similarly and that the accelerometer-only and combined non-weighted methods performed significantly worse. As predicted by the results concerning identifiability of the joint axis in \cite{Nowka2019_ECC}, the accelerometer only allows identifiability up to sign pairing in the case where the joint axis is horizontal. A possible explanation for the worse results for $\hat{j}_1$ could be that the vertical acceleration component is much larger than the horizontal acceleration component (gravitational acceleration $g$ compared to the horizontal acceleration induced by flexion/extension of the joint). This might cause the scalar product between the acceleration vectors and horizontal vectors other than the joint axis, to be close to zero, causing ambiguity in the determination of a global minimum of the cost function.

The results have shown that, out of the compared methods, the proposed method is the only one that consistently achieves the lowest angular error for both $j_1$ and $j_2$ for all motions considered. Even in the horizontal axis case, where the accelerometer do not provide information about the sign pairing, the proposed method has at least similar performance to that of the gyroscope-only method. However, these results should encourage the user to not restrict the joint axis to the horizontal plane while collecting the estimation data, when it is possible. 

\subsection{Choice of weighting variables}
The proposed weighting serves two purposes. First, the gyroscope weight $w_\omega(k)$ scales the gyroscope residuals to make the variance match that of the accelerometer residual. For our sensors we identified the standard deviations of the measurement noise to be $\sigma_\omega = \SI{0.0050}{\radian\per\second}$ for the gyroscope and $\sigma_a = \SI{0.0346}{\metre\per\second\squared}$ for the accelerometer. Which means that $w_\omega(k) = w_0 = \frac{\sigma_a}{\sigma_\omega} = \SI{6.92}{\metre\per\radian\per\second}$. Therefore, the gyroscope residuals are given more weight after this constant scaling, which prevents the accelerometer residuals from having a significantly larger impact on the value of the cost function. The second purpose is to give less weight to accelerometer residuals that violate the approximation of the constraint $\eqref{eq:acc_constrant_nonzero}$. Therefore, we chose to include a penalty term that depend on the difference in magnitude of the two accelerometer measurements in \eqref{eq:acc_weight}. However, a potential weakness of the proposed weighting strategy is that some measurements that violate the acceleration constraint might still get a weight $w_a(k) = 1$, if the rotational acceleration components of both sensors are the same. Another approach that was considered was to select a penalty term to match the order of magnitude of $j_1^\top K_1 r_1 - j_2^\top K_2 r_2$, since this is the actual approximation error of the acceleration constraint. However, this approach assumes prior knowledge of the position of each sensor with respect to the joint axis (the vectors $r_i$). It is known that identification of $r_i$ is ambiguous for hinge-joints \cite{olsson2017experimental}, since in that case any vector that points from the origin of the sensor frame to any point on the infinitely extended joint axis satisfies the kinematic constraints in \eqref{eq:a1}. Therefore, the chosen approach was deemed less restrictive. The proposed weighting is free from tuning parameters that need to be manually selected by the user. It depends only on the noise standard deviations, which are typically computed as part of IMU calibration, and the measurements themselves.

\section{Conclusions and future work} \label{sec:conclusions}
A novel method for estimating the joint axis of a hinge joint in a kinematic chain has been proposed. The method uses a weighted combination of accelerometer and gyroscope readings obtained from two 6-DOF IMUs, one attached to each segment of the hinge joint. The proposed method was compared to methods which use gyroscope- or accelerometer information alone, or combined but non-weighed. The methods were evaluated using real data collected from sensors strapped to a rigid upper limb exoskeleton frame. The data consisted of nine different data sets, with different combinations of fast, slow and mixed movement speed under conditions where the joint axis was either free to move or constrained to a horizontal or vertical orientation. In contrast to all other methods, the proposed method yields accurate joint axis estimates for all movement speeds, under the free- and vertical joint axis conditions. For the horizontal joint axis condition, the proposed method performs similarly to the gyroscope-only method when sign pairing cannot be determined by the accelerometer.

The proposed method can be used in any system with approximate hinge joints without the need for accurate sensor placement or precise calibration motions. The inclusion of the weighted accelerometer information overcomes many limitations of the previously proposed methods. Therefore, the proposed method is an important step towards full plug-and-play motion tracking, where arbitrarily placed sensors calibrate themselves with minimal interference required by the user.

Not included in the present study is an evaluation on human data, where soft tissue artifacts may be present and cause the rigid body assumption to be violated. We have recently collected data from human movement of knee- and finger joints, which will be used in further evaluation of the method. Another interesting extension of the method would be to combine joint axis estimation with joint center estimation.

\bibliographystyle{IEEEtran}
\bibliography{IEEEabrv,myrefs}

\begin{thebibliography}{10}
\providecommand{\url}[1]{#1}
\csname url@samestyle\endcsname
\providecommand{\newblock}{\relax}
\providecommand{\bibinfo}[2]{#2}
\providecommand{\BIBentrySTDinterwordspacing}{\spaceskip=0pt\relax}
\providecommand{\BIBentryALTinterwordstretchfactor}{4}
\providecommand{\BIBentryALTinterwordspacing}{\spaceskip=\fontdimen2\font plus
\BIBentryALTinterwordstretchfactor\fontdimen3\font minus
  \fontdimen4\font\relax}
\providecommand{\BIBforeignlanguage}[2]{{%
\expandafter\ifx\csname l@#1\endcsname\relax
\typeout{** WARNING: IEEEtran.bst: No hyphenation pattern has been}%
\typeout{** loaded for the language `#1'. Using the pattern for}%
\typeout{** the default language instead.}%
\else
\language=\csname l@#1\endcsname
\fi
#2}}
\providecommand{\BIBdecl}{\relax}
\BIBdecl

\bibitem{ahmad2013reviews}
N.~Ahmad, R.~A.~R. Ghazilla, N.~M. Khairi, and V.~Kasi, ``Reviews on various
  inertial measurement unit (imu) sensor applications,'' \emph{International
  Journal of Signal Processing Systems}, vol.~1, no.~2, pp. 256--262, 2013.

\bibitem{kok2017using}
M.~Kok, J.~D. Hol, and T.~B. Sch{\"o}n, ``Using inertial sensors for position
  and orientation estimation,'' \emph{Foundations and Trends® in Signal
  Processing}, vol.~11, no. 1-2, pp. 1--153, 2017.

\bibitem{kok2015indoor}
------, ``Indoor positioning using ultrawideband and inertial measurements,''
  \emph{IEEE Transactions on Vehicular Technology}, vol.~64, no.~4, pp.
  1293--1303, 2015.

\bibitem{5504750}
A.~D. {Young}, ``Use of body model constraints to improve accuracy of inertial
  motion capture,'' in \emph{2010 International Conference on Body Sensor
  Networks}, June 2010, pp. 180--186.

\bibitem{Laidig2017_ICORR}
D.~Laidig, T.~Schauer, and T.~Seel, ``Exploiting kinematic constraints to
  compensate magnetic disturbances when calculating joint angles of approximate
  hinge joints from orientation estimates of inertial sensors,'' in \emph{Proc.
  of 15th IEEE Conference on Rehabilitation Robotics (ICORR)}, London, UK,
  2017, p. 971–976.

\bibitem{SeelSchauerRaisch2012}
T.~Seel, T.~Schauer, and J.~Raisch, ``Joint axis and position estimation from
  inertial measurement data by exploiting kinematic constraints,'' in
  \emph{IEEE International Conference on Control Applications}, 2012, pp.
  45--49.

\bibitem{CraboluPaniCereatti2016}
M.~Crabolu, D.~Pani, and A.~Cereatti, ``Evaluation of the accuracy in the
  determination of the center of rotation by magneto-inertial sensors,'' in
  \emph{Sensors Applications Symposium IEEE}, 2016.

\bibitem{olsson2017experimental}
F.~Olsson and K.~Halvorsen, ``Experimental evaluation of joint position
  estimation using inertial sensors,'' in \emph{Information Fusion (Fusion),
  2017 20th International Conference on}.\hskip 1em plus 0.5em minus
  0.4em\relax IEEE, 2017, pp. 1--8.

\bibitem{s18061882}
T.~McGrath, R.~Fineman, and L.~Stirling, ``An auto-calibrating knee
  flexion-extension axis estimator using principal component analysis with
  inertial sensors,'' \emph{Sensors}, vol.~18, no.~6, 2018.

\bibitem{laidig_2017_2d}
D.~Laidig, P.~M\"{u}ller, and T.~Seel, ``Automatic anatomical calibration for
  {IMU}-based elbow angle measurement in disturbed magnetic fields,''
  \emph{Current Directions in Biomedical Engineering}, vol. 3(2), pp. 167--170,
  2017.

\bibitem{jarrasse2012connecting}
N.~Jarrasse and G.~Morel, ``Connecting a human limb to an exoskeleton,''
  \emph{IEEE Transactions on Robotics}, vol.~28, no.~3, pp. 697--709, 2012.

\bibitem{Nowka2019_ECC}
\BIBentryALTinterwordspacing
D.~Nowka, M.~Kok, and T.~Seel, ``On motions that allow for identification of
  hinge joint axes from kinematic constraints and 6d imu data,'' in
  \emph{accepted for European Control Conference (ECC)}, Nice, France, 2019.
  [Online]. Available:
  \url{https://www.control.tu-berlin.de/wiki/images/b/b3/Nowka2019_ECC.pdf}
\BIBentrySTDinterwordspacing

\bibitem{kuderle2018increasing}
A.~K{\"u}derle, S.~Becker, and C.~Disselhorst-Klug, ``Increasing the robustness
  of the automatic imu calibration for lower extremity motion analysis,''
  \emph{Current Directions in Biomedical Engineering}, vol.~4, no.~1, pp.
  439--442, 2018.

\bibitem{nocedalopt}
S.~Wright and J.~Nocedal, \emph{Numerical Optimization}, 2nd~ed.\hskip 1em plus
  0.5em minus 0.4em\relax Springer Series in Operations Research, 2006.

\bibitem{Xsens2017}
Xsens, ``{MTw Awinda Wireless Motion Tracker},'' (Online)
  \url{https://www.xsens.com/download/pdf/documentation/mtw2-awinda/MTw2-Awinda.pdf}
  accessed December 2017, 2017.

\end{thebibliography}

\end{document}